\documentclass{emulateapj}

\usepackage{apjfonts}
\usepackage{amsmath}
\newcommand{\etal}{et al.}

\newcommand{\CIV}{C{\sevenrm IV}\ }

\newcommand{\MgII}{Mg{\sevenrm II}\ }

\newcommand{\bracket}[1]{\left\langle#1\right\rangle}
 
 \font\sevenrm=cmr7 scaled 1000

\slugcomment{accepted for publication in ApJ}

\begin{document}
\title{Do broad absorption line quasars live in different environments from ordinary quasars?}

\shorttitle{BAL QUASAR CLUSTERING}

\shortauthors{SHEN ET AL.}

\author{Yue Shen\altaffilmark{1}, Michael A. Strauss\altaffilmark{1}, Patrick B. Hall\altaffilmark{2},
Donald P. Schneider\altaffilmark{3}, Donald G.
York\altaffilmark{4}, Neta A. Bahcall\altaffilmark{1}}

\altaffiltext{1}{Princeton University Observatory, Princeton, NJ
08544.}

\altaffiltext{2}{Dept. of Physics \& Astronomy, York University,
4700 Keele St., Toronto, ON, M3J 1P3, Canada. }

\altaffiltext{3}{Department of Astronomy and Astrophysics, 525
Davey Laboratory, Pennsylvania State University, University Park,
PA 16802.}

\altaffiltext{4}{University of Chicago and Enrico Fermi Institute,
5640 So. Ellis Avenue, Chicago, IL 60637}

\begin{abstract}
We select a sample of $\sim 4200$ traditionally defined broad
absorption line quasars (BALQs) from the Fifth Data Release quasar
catalog of the Sloan Digital Sky Survey. For a statistically
homogeneous quasar sample with $1.7\le z\le 4.2$, the BAL quasar
fraction is $\sim 14\%$ and is almost constant with redshift. We
measure the auto-correlation of non-BAL quasars (nonBALQs) and the
cross-correlation of BALQs with nonBALQs using this statistically
homogeneous sample, both in redshift space and using the projected
correlation function. We find no significant difference between
the clustering strengths of BALQs and nonBALQs. Assuming a
power-law model for the real space correlation function
$\xi(r)=(r/r_0)^{-1.8}$, the correlation length for nonBALQs is
$r_0=7.6\pm 0.8\ h^{-1}{\rm Mpc}$; for BALQs, the
cross-correlation length is $r_0=7.4\pm 1.1\ h^{-1}{\rm Mpc}$. Our
clustering results suggest that BALQs live in similar large-scale
environments as do nonBALQs.
\end{abstract}

\section{INTRODUCTION}
Broad absorption line (BAL) quasars constitute a significant
fraction, $\sim 10-20\%$ of the entire quasar population (Weymann
2002; Tolea et al. 2002; Hewett \& Foltz 2003; Reichard et al.
2003a; Trump et al. 2006). While the broad absorption features are
usually explained by invoking an outflowing wind, it remains
unclear how special BAL outflows are. Is it that $\sim 10-20\%$ of
the whole quasar population has such BAL outflows, or do most
quasars have BAL outflows, but with an average wind covering
fraction of $\sim 10-20\%$? The former case would suggest that
BALQs are an intrinsically physically distinct class of quasars,
while the latter model implies that the BAL phenomenon is an
accident of orientation. Current observations are consistent with
most BALQs being intrinsically no different from ordinary quasars,
with the BAL phenomenon arising when the line-of-sight cuts
through low covering fraction outflows, as produced in some disk
wind models (e.g., Murray et al. 1995; Proga et al. 2000; Elvis
2000 and references therein). Spectropolarimetry shows that there
are absorption-line-free lines of sight in BAL quasars, and the
wind covering fraction is inferred to be low from constraints on
ultraviolet emission-line scattering (e.g., Hamann et al. 1993;
Ogle et al. 1999; but cf. Brotherton et al. 2006). The continuum
and emission line properties of BALQs suggest a disk
wind/orientational-obscuration geometry (e.g., Reichard et al.
2003a,b). Gallagher et al. (2007) found no compelling evidence for
inherent differences in composite mid-infrared through X-ray SEDs
between BAL and non-BAL quasars of comparable luminosity. Shen et
al. (2007b) found that the distribution of black hole masses
measured from the \MgII virial estimator is almost identical
between a \CIV BAL and a non-BAL quasar sample matched in redshift
and luminosity, similar to what was found in Ganguly et al.
(2007).

Of course, there are still many unsettled issues regarding BALQs.
For example, the exact geometry of these outflows is still under
theoretical investigation (e.g., Proga 2007) and observational
debate.  It has been argued based on the properties of radio-loud
BAL quasars (Becker et al. 2000, and references therein) that the
outflow cannot be purely equatorial in at least some cases. For
example, Brotherton et al. (2006) reported a radio-loud BAL quasar
whose spectropolarimetry is consistent with it being viewed closer
to pole-on than edge-on.  Based on radio variability arguments
which yield high brightness temperatures, Zhou et al. (2006) and
Ghosh \& Punsly (2007) argue that some BAL quasars are viewed
nearly along the polar axis (but see Blundell \& Kuncic 2007).
However, the exact geometry of BAL outflows is immaterial when
considering whether or not BAL quasars form a distinct
subpopulation of quasars.

A somewhat different test of whether BALQs are a distinct quasar
subpopulation is to search for differences in their large-scale
environments as probed by their clustering properties. Since
quasars live in dark matter halos, and more massive dark matter
halos are more highly biased and therefore have intrinsically
stronger spatial clustering, the clustering properties of quasars
can constrain the masses of their host dark matter halos (Shen et
al. 2007a and references therein). A difference in the clustering
properties between BALQs and nonBALQs will directly imply BALQs
belong to a distinct class, hence such a test is crucial to the
current BAL quasar picture. However, such analyses require large
and statistically homogeneous BAL quasar samples, and were thus
impractical in earlier studies on BALQs (e.g., Weymann et al.
1991; Hewett \& Foltz 2003). The Sloan Digital Sky Survey (SDSS,
York et al. 2000) has made such studies possible by providing a
large sample of optically-selected quasars (see Schneider et al.
2007 for the latest SDSS quasar catalog), from which we can
construct homogeneous subsamples for clustering studies. BAL
quasar catalogs constructed using SDSS quasars have greatly
increased the number of BALQs known in the literature. The latest
SDSS BAL quasar catalog, based on Data Release Three (Abazajian et
al. 2005), was compiled by Trump et al. (2006) and contains $\sim
2000$ traditional BALQs.

In this paper, we extend the traditional BAL quasar catalog to the
fifth SDSS data release (DR5, Adelman-McCarthy et al. 2007), and
investigate the clustering properties of BALQs based on
statistically homogeneous samples. We describe the construction of
our BAL quasar samples in \S\ref{sec:sample}, present the
clustering measurements in \S\ref{sec:clustering} and discuss our
results in \S\ref{sec:disc}. Throughout this paper we adopt a flat
$\Lambda$CDM cosmology: $\Omega_M=0.26$, $\Omega_\Lambda=0.74$ and
$h=0.71$ (Spergel \etal\ 2007).

\section{Sample Selection}\label{sec:sample}
Our parent sample is the published DR5 quasar catalog (Schneider
et al. 2007), which contains 77,429 quasars. About half the
quasars in this catalog were targeted using a uniform algorithm
(Richards et al. 2002) and will be used to construct statistically
homogeneous subsamples for our clustering analysis (e.g., Richards
et al. 2006; Shen et al. 2007a).

BALQs are identified using the traditional ``Balnicity Index''
(BI) criterion (Weymann et al. 1991, see the definition in their
Appendix A). Our procedure is similar to Reichard et al. (2003a)
and Trump et al. (2006): the input spectrum is fitted by a
composite quasar spectrum with adjustable power-law scaling and
dust reddening, normalized using different continuum windows (see
section 3.3 in Reichard et al. 2003a for details). We have used
the Vanden Berk et al. (2001) composite quasar spectrum
constructed from the SDSS Early Data Release (EDR, Stoughton et
al. 2002), and the Pei (1992) SMC extinction curve (e.g., Hopkins
et al. 2004). The continuum normalization windows are $1725\pm25$
\AA\ for \CIV and $3150\pm 25$ \AA\ ($0.5\le z\le1.9$) and
$2200\pm25$ \AA\ ($1.9< z\le2.1$) for \MgII. Once we have the
composite fits, BALQs are identified for quasars with \CIV
absorption at least $2000\ {\rm km\ s^{-1}}$ broad located between
3000 and 25,000 ${\rm km\ s^{-1}}$ blueward of the quasar
redshift; or with \MgII absorption at least $1000\ {\rm km\
s^{-1}}$ broad located between 0 and 25,000 ${\rm km\ s^{-1}}$
blueward of the quasar redshift. Thus given the SDSS spectral
coverage, complete BAL quasar samples can only be identified
within $1.7\le z\le 4.2$ for \CIV and $0.5\le z\le 2.1$ for \MgII,
and so we restrict ourselves to these redshift ranges.

\begin{figure*}
  \centering
    \includegraphics[width=0.85\textwidth]{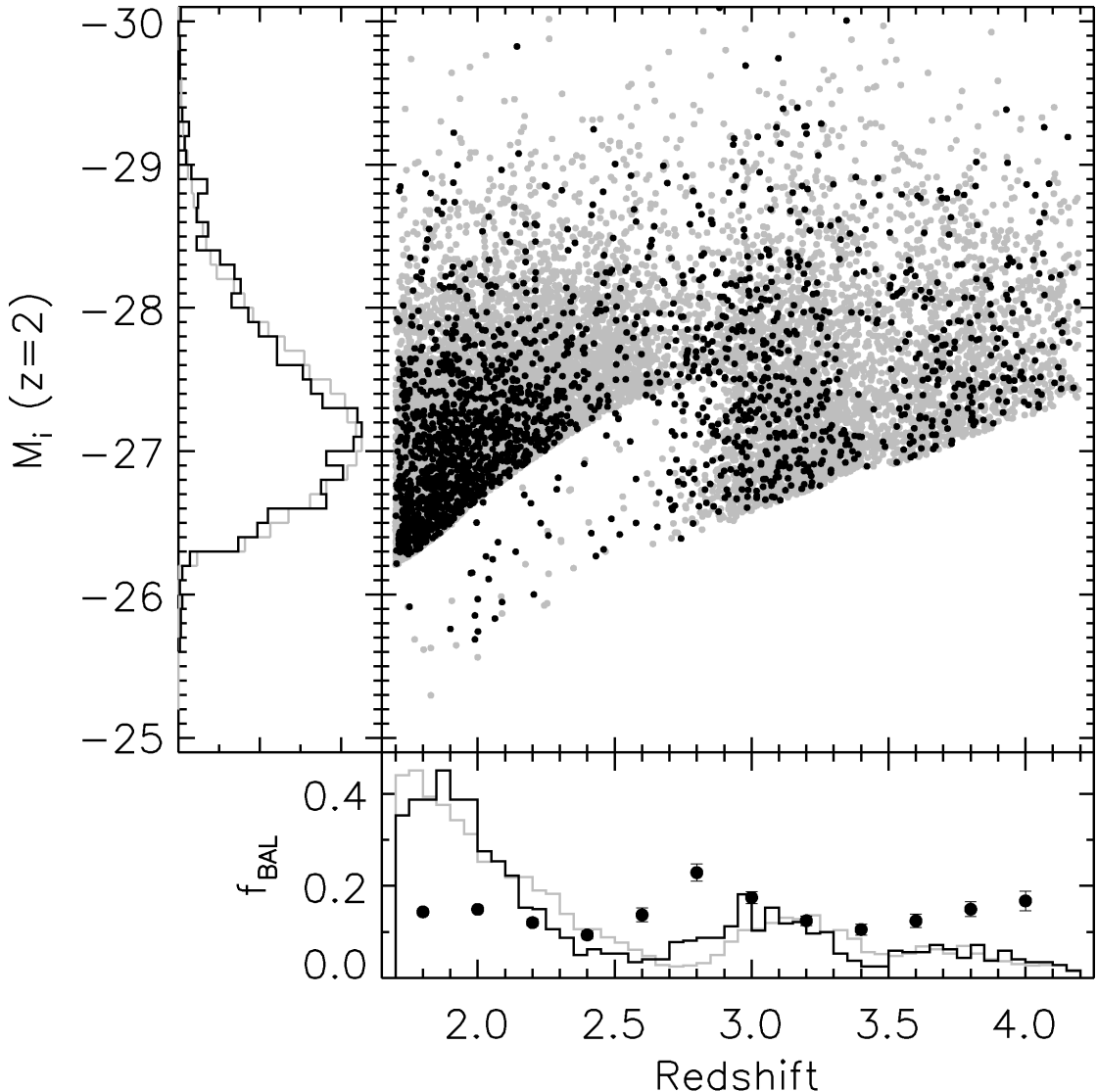}
    \caption{Distribution of our statistically homogeneous samples of
    \CIV BALQs (black dots) and nonBALQs (gray dots)
    in the redshift-luminosity plane.  The left and bottom panels show the
    histograms of $i-$band absolute magnitudes ($K-$corrected to $z=2$,
    Richards et al. 2006) and redshifts, gray for nonBALQs and black for BALQs. In the
    bottom panel we also show the BAL quasar fraction (with statistical errors) as function of
    redshift in filled circles. Despite the jump around $z\sim 2.7$, which may be caused by the
    general inefficiency in color selection around this redshift (see text), the BAL quasar fraction
    $\sim 14\%$ is almost constant with redshift. }
    \label{fig:dist}
\end{figure*}

\begin{figure*}
  \centering
    \includegraphics[width=0.45\textwidth]{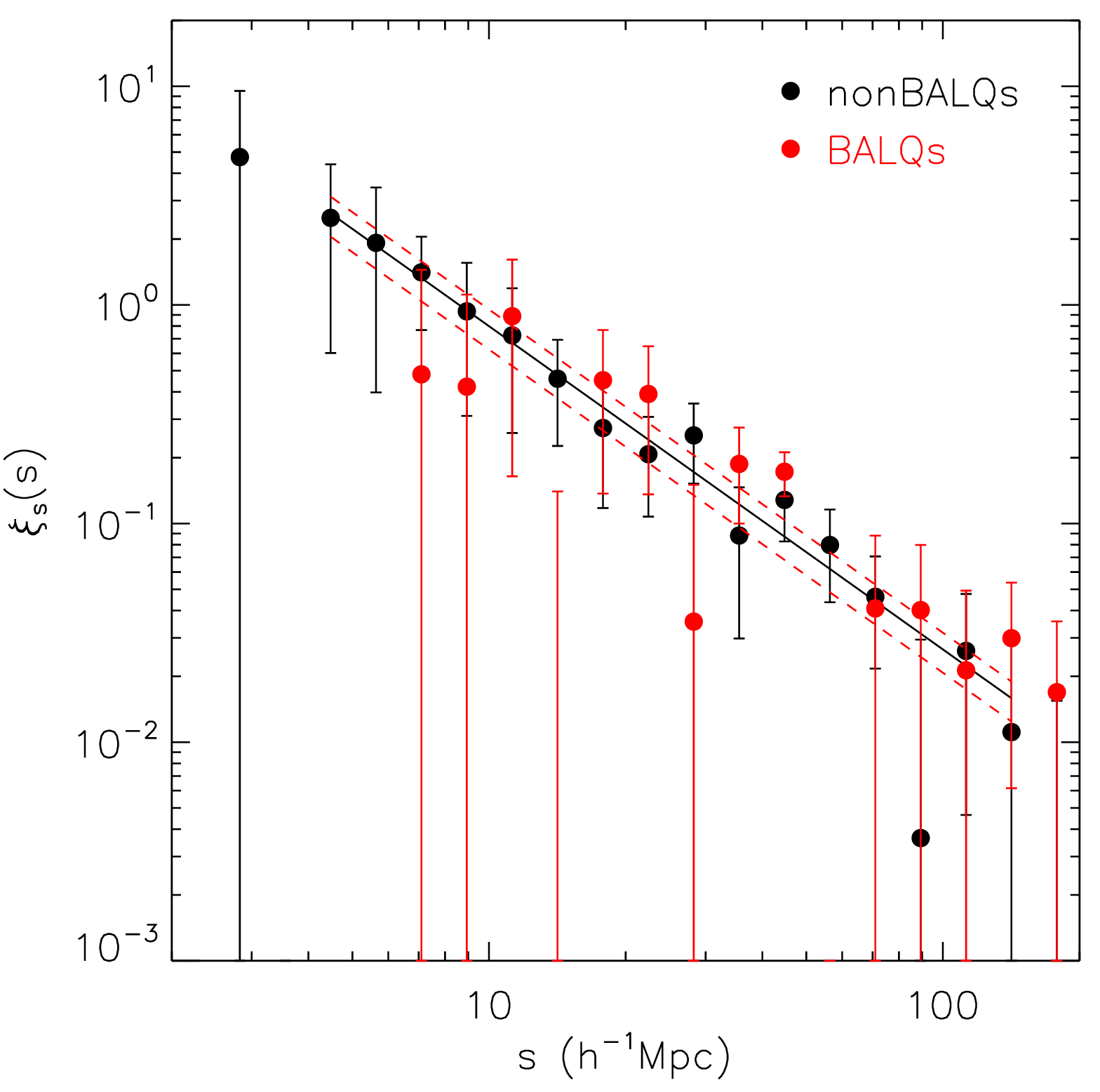}
    \includegraphics[width=0.45\textwidth]{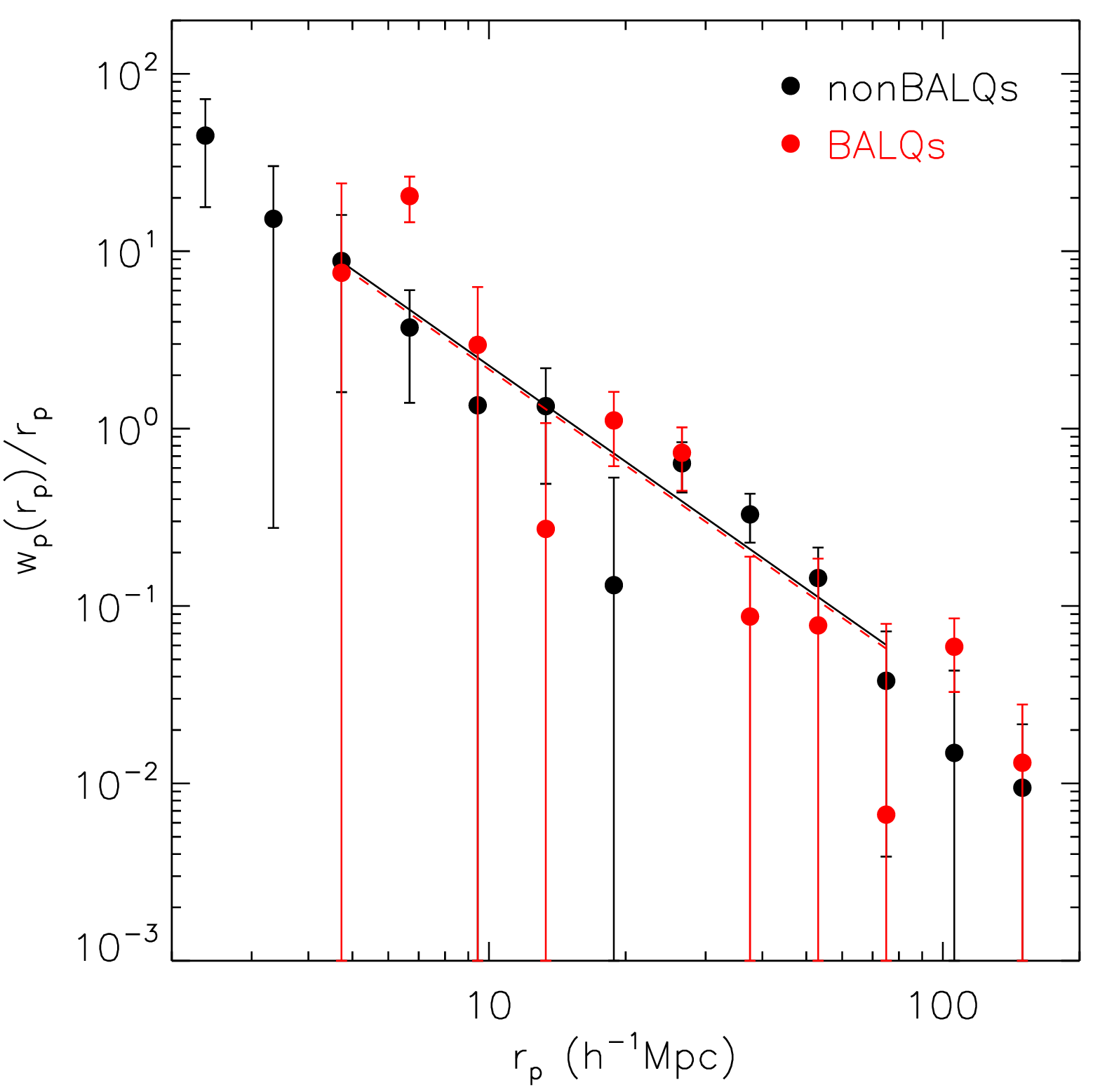}
    \caption{Redshift space (left) and projected (right) correlation functions.
    Black
    and red colors denote the nonBAL quasar auto-correlation and the
    BAL-nonBAL quasar cross-correlation, respectively. Errors are
    estimated using the jackknife method. The solid and dashed lines are fitted
    power-law models. The fitting ranges are $4<s<150\ h^{-1}{\rm Mpc}$ for
    the redshift space correlation function, and $4<r_p<80\ h^{-1}{\rm Mpc}$ for the
    projected correlation function. The upper and lower dashed lines in the left panel
    are two fits to the BAL-nonBAL quasar cross correlation, including and
    excluding the two negative data points at $s\approx 15$ and $55\
    h^{-1}{\rm Mpc}$. }
    \label{fig:CF}
\end{figure*}

The procedure we describe in this paper is not identical to that
of Trump et al. (2006).  In particular, when measuring absorption
troughs we have boxcar smoothed the input spectrum by 15 pixels,
while Trump et al. used a smoothing window of three pixels. This
increases our sensitivity to weak BAL features but also greatly
increases the rate of false detections. Also, we are using the
Vanden Berk et al. (2001) composite spectrum for all input
spectra, instead of using different templates as Trump et al.
(2006) did, which allow for variations in emission-line strength
and shape.

Although automatic pipelines are efficient for large samples,
there will inevitably be false and missing identifications. For
this reason, we examined the spectra of the entire DR5 quasar
catalog by eye, and moved objects into or out of the BAL quasar
catalog. We found that out automated procedure missed only about
$1\%$ of BALQs in our final sample; while the false detection rate
is quite high (over half are actually quasars with high velocity
narrow absorption lines), primarily caused by the 15-pixel boxcar
smoothing. Therefore our manual inspection is necessary to include
only BALQs in our sample. In comparison to the DR3 portion in the
Trump et al. (2006) catalog, some weak BALQs are not included in
our catalog, while we also include some strong BALQs that are
apparently missing in the Trump et al. catalog. Ganguly et al.
(2007) manually inspected the Trump et al. catalog, and also
concluded that automatic ID procedures are not perfect. Our final
catalog contains 4203 BALQs. This list is perhaps still incomplete
to some extent, and less restrictive BAL criteria such as the AI
index (Hall et al. 2002) used in the Trump et al. catalog will
certainly increase the number of BALQs. The distribution of the
\CIV balnicity index is similar to what was found by Tolea et al.
(2002) and Reichard et al. (2003a), with more BALQs at the lower
BI end. But determining the exact distribution of the BI index
will require more careful work on individual spectral fits, which
we defer to a BAL quasar catalog paper in preparation.

Quasars showing \MgII BAL almost always show \CIV BAL as well, but
the reverse is not true. Hence our BAL quasar catalog is close to
complete for $1.7\le z\le4.2$, where the spectral coverage allows
us to identify \CIV BALQs. We now further restrict ourselves to
uniformly-selected quasars based on color-selection (Richards et
al. 2002). These uniformly-targeted quasars are flux limited to
$i=19.1$ at $z\lesssim 3$ and $i=20.2$ at $z\gtrsim 3$,
\footnote{There are a few $i>19.1$ quasars at $z\lesssim 3$ which
were selected by the high-z ($griz$) branch of the targeting
algorithm (Richards et al. 2002). The fraction of these objects is
tiny ($\lesssim 2\%$) and does not affect our analysis.} and are
selected using the final quasar target algorithm (see details in
Richards et al. 2002) implemented after DR1 (Abazajian et al.
2003). This results in statistically homogeneous samples of 12,117
nonBALQs and 1942 BALQs in the redshift range $1.7\le z\le 4.2$.
The distributions of these nonBALQs and BALQs in the
redshift-luminosity diagram are shown in Fig. \ref{fig:dist} as
gray and black dots respectively. Their distributions are almost
indistinguishable, and the BAL quasar fraction\footnote{The BAL
quasar fraction here is the raw fraction, i.e., without
corrections for intrinsic extinction. Dai et al. (2007) recently
measured an intrinsic BAL fraction (using the traditional BI
definition) of $\sim 20\%$ based on a sample of 2MASS (Skrutskie
et al. 2006) matched SDSS quasars (see also Hewett \& Foltz
2003).}, $\sim 14\%$, is nearly constant with redshift. There is
an apparent excess of BALQs at $z\sim 2.7$. This is because around
this redshift the colors of quasars are similar to those of F
stars (Fan 1999) and the quasar target selection becomes less
efficient; BALQs have different broad band colors and are perhaps
less sensitive to this selection inefficiency. Despite this
detail, there is little evidence that the BAL quasar fraction
changes significantly within this redshift range. We take these
uniformly-selected subsamples as our clustering subsamples. Given
the similarity in the redshift and luminosity distributions of
BALQs and nonBALQs in these subsamples, we can fairly compare the
difference, if there is any, in the clustering properties of BALQs
and nonBALQs.

The complete list of BALQs and nonBALQs with flags indicating
whether or not each is included in the clustering analysis is
presented in Table 1 of Shen et al. (2007b).

\section{clustering analysis}\label{sec:clustering}
Following Shen et al. (2007a), we generate random catalogs
according to the detailed angular and radial geometries of our
clustering subsamples. We measure the auto-correlation of
nonBALQs, and the cross-correlation of nonBALQs around BALQs. Our
BAL quasar sample is quite sparse, and therefore its
auto-correlation function is too noisy to be useful. We use the
cross-correlation technique to boost the clustering signal and to
achieve reasonable measurements. While it is straightforward to
measure the redshift space correlation function $\xi_s(s)$, such
measurements suffer from redshift distortions and the
uncertainties in redshift determinations. Using the projected
correlation function $w_p(r_p)$ avoids these problems (e.g., Davis
\& Peebles 1983), and gives an unbiased estimate of the real space
correlation function $\xi(r)$. In estimating errors, we use
jackknife resampling following Shen et al. (2007a).

For the auto-correlation function of nonBALQs, we use the Landy \&
Szalay (1993) estimator:
\begin{equation}
\xi_s(s),\
\xi_s(r_p,\pi)=\frac{\bracket{DD}-2\bracket{DR}+\bracket{RR}}{\bracket{RR}}\
, \label{eq:landy_szalay}
\end{equation}
where $\bracket{DD}$, $\bracket{DR}$, and $\bracket{RR}$ are the
normalized numbers of data-data, data-random and random-random
pairs, respectively, in our desired bins.  The projected correlation
function is then obtained by integrating the two-dimensional
redshift space correlation function $\xi_s(r_p, \pi)$ along the
line-of-sight ($\pi$) direction:
\begin{equation}
w_p(r_p)=2\int_0^\infty d\pi\,\xi_s(r_p,\pi)\ .
\end{equation}
In practice we cut this integral at $\pi_{\rm cutoff}=50\
h^{-1}{\rm Mpc}$. We find $\pi_{\rm cutoff}=70\ h^{-1}{\rm  Mpc}$
and $100 \ h^{-1}{\rm  Mpc}$ give essentially identical results,
but with larger uncertainties since more noise is added.

For the cross-correlation of nonBALQs around BALQs, we use the
estimator (e.g., Coil et al. 2007)
\begin{equation}\label{eqn:cross_CF}
\xi_s(s),\ \xi_s(r_p,\pi)=\frac{\bracket{BN}}{\bracket{BR}}-1\ ,
\end{equation}
where $\bracket{BN}$ and $\bracket{BR}$ are normalized BAL-nonBAL
quasar and BAL quasar-random pairs at a given separation. To
compute the projected cross-correlation function, we again use
$\pi_{\rm cutoff}=50\ h^{-1}{\rm Mpc}$.

Our results are shown in Fig. \ref{fig:CF} for the redshift space
correlation function (left) and projected correlation function
(right), along with fitted power-law models. For the redshift
space correlation function, the fitted power-law
$\xi_s(s)=(s/s_0)^{-\delta}$ has $s_0=8.6\pm 1.3\ h^{-1}{\rm
Mpc}$, $\delta=1.5\pm 0.2$ for nonBALQs for the fitting range
$4<s<150\ h^{-1}{\rm Mpc}$; $s_0=7.3\pm 1.3\ h^{-1}{\rm Mpc}$ or
$9.7\pm 1.3\ h^{-1}{\rm Mpc}$ for the nonBAL-BAL quasar cross
correlation when we include all the data points or exclude the two
negative points at $s\approx 15$ and $55\ h^{-1}{\rm Mpc}$
respectively, for the same fitting range $4<s<150\ h^{-1}{\rm
Mpc}$ and fixing the power-law slope to be the same as the nonBAL
quasar case. For the projected correlation function and assuming a
power-law real space correlation function $\xi(r)=(r/r_0)^{-1.8}$
we have: $r_0=7.6\pm 0.8\ h^{-1}{\rm Mpc}$ and $r_0=7.4\pm 1.1\
h^{-1}{\rm Mpc}$ for the nonBAL quasar and cross-correlation cases
respectively, for the fitting range $4<r_p<80\ h^{-1}{\rm Mpc}$.
Although the data points scatter around these fitted power-law
models (especially for the nonBAL-BAL quasar cross correlation),
and the jackknife error estimator might be inadequate for sparse
samples, we find no convincing evidence that the auto-correlation
function of nonBALQs is different from their cross-correlation
function with BALQs. Hence our results suggest that BALQs and
nonBALQs live in similar environments on large scales, and
therefore lie in similar dark matter halos.

The quasar correlation function evolves with redshift, especially
for $z\gtrsim 2.8$ (Shen et al. 2007a). Hence we divide our sample
into two redshift bins, $1.7\le z\le 2.8$ and $2.8\le z\le 4.2$,
and repeat the clustering analysis. We still did not find any
compelling difference between the auto-correlation function of
nonBALQs and the cross-correlation function of nonBALQs-BALQs
within each redshift bin, although the cross-correlation function
is quite noisy for the higher redshift bin due to the smaller
numbers of both nonBALQs and BALQs.

A potential drawback of our analysis is that the velocity width
cut (2000 ${\rm km\ s^{-1}}$ for C{\sevenrm IV}) used in the BI
definition of a BAL quasar is rather arbitrary. By imposing such a
cut we are rejecting high-velocity narrow absorption line (NAL)
quasars into the ``nonBAL'' quasar sample. The fraction of these
NAL quasars to the entire population is perhaps $\gtrsim 10\%$
based on the excess fraction of BALQs selected using the AI
definition (which loosens the velocity width cut to be 1000 ${\rm
km\ s^{-1}}$) in the Trump et al. catalog. In addition, $\sim
25\%$ of all quasars show associated absorption line (AAL) systems
(e.g., Ganguly et al. 2001; Vestergaard 2003; Ganguly \&
Brotherton 2007), which are not included in our BI based BAL
quasar sample. NAL and AAL quasars together make up $\sim 37\%$ of
the entire quasar population and $\sim 40\%$ of our nonBAL quasar
sample based on the estimations by Ganguly et al. (2007). Of
course, not all of these NAL and AAL systems are intrinsically
associated with the quasar; some must be intervening systems. If
those intrinsic NAL and AAL systems\footnote{A recent study on
$\sim 400$ \MgII AAL systems suggests that the absorption
originates from gas in the host galaxies of quasars (e.g., Vanden
Berk et al., 2007), suggesting that they would have the same
clustering properties as ordinary quasars. A clustering analysis
of AAL systems using a large sample from SDSS DR5 is underway to
test this hypothesis.} have the same clustering as BALQs but have
different clustering from absorption-free quasars, then including
them in our nonBAL quasar clustering subsample would potentially
dilute any possible difference in clustering between BAL/NAL/AAL
quasars and absorption-free quasars. To test this, we have
selected a ``cleaned'' sample of $\sim 8000$ absorption-free
quasars from our nonBAL quasar clustering subsample, and measured
their auto-correlation function. Although the signal-to-noise
ratio is lower, we found no convincing difference from using the
whole nonBAL quasar sample, reassuring that we are not biasing our
results. This result is also consistent with the general picture
that quasar outflows are ubiquitous, while the modest covering
factor of these various (BAL or NAL) outflows causes the different
appearances of quasars with and without absorption features.

\section{Discussion}\label{sec:disc}
We have made the first attempt to measure the clustering
properties of broad absorption quasars, based on the largest and
most homogeneous BAL quasar sample available. However, the
sparseness of the BAL quasar sample ($\sim 0.5$ per ${\rm deg}^2$
over a wide redshift range) prohibits direct measurement of the
BAL quasar auto-correlation function. To boost the clustering
signal, we have cross-correlated BALQs with the $\sim 6$ times
larger sample of nonBALQs at the same redshifts. Our results
suggest that BALQs have similar clustering properties as nonBALQs
on the scales probed by our sample, and hence should reside in
similar large-scale environments.

Our clustering results provide support to the idea that BALQs are
drawn uniformly from the overall underlying population of quasars,
as would be the case if (for example) BAL troughs arose in disk
wind outflows (Murray et al. 1995; Proga et al. 2000; Elvis 2000).
The nearly constant BAL quasar fraction with redshift in our
optical quasar sample is also consistent with this picture. It may
still be that members of the rare subclass of FeLoBALQs are
intrinsically different from ordinary quasars (Farrah et al.
2007), and may reside in different environments. Unfortunately,
the small number of FeLoBALQs in our sample makes a clustering
analysis of them too noisy to be useful.

\acknowledgements We thank Linhua Jiang for pointing out some
missing objects in our original BAL quasar sample and Jenny Greene
for helpful discussions. YS and MAS acknowledge the support of NSF
grants AST-0307409 and AST-0707266. PBH is supported by NSERC. DPS
acknowledges the support of NSF grant AST-0607634.

Funding for the SDSS and SDSS-II has been provided by the Alfred
P. Sloan Foundation, the Participating Institutions, the National
Science Foundation, the U.S. Department of Energy, the National
Aeronautics and Space Administration, the Japanese Monbukagakusho,
the Max Planck Society, and the Higher Education Funding Council
for England. The SDSS Web Site is http://www.sdss.org/.

The SDSS is managed by the Astrophysical Research Consortium for
the Participating Institutions. The Participating Institutions are
the American Museum of Natural History, Astrophysical Institute
Potsdam, University of Basel, University of Cambridge, Case
Western Reserve University, University of Chicago, Drexel
University, Fermilab, the Institute for Advanced Study, the Japan
Participation Group, Johns Hopkins University, the Joint Institute
for Nuclear Astrophysics, the Kavli Institute for Particle
Astrophysics and Cosmology, the Korean Scientist Group, the
Chinese Academy of Sciences (LAMOST), Los Alamos National
Laboratory, the Max-Planck-Institute for Astronomy (MPIA), the
Max-Planck-Institute for Astrophysics (MPA), New Mexico State
University, Ohio State University, University of Pittsburgh,
University of Portsmouth, Princeton University, the United States
Naval Observatory, and the University of Washington.

Facilities: Sloan

\end{document}